\documentclass[sigconf,screen]{acmart}

\AtBeginDocument{%
  }

\setcopyright{acmlicensed}
\copyrightyear{2018}
\acmYear{2018}
\acmDOI{XXXXXXX.XXXXXXX}
\acmConference[]{}{}{}

\acmBooktitle{}
\acmISBN{978-1-4503-XXXX-X/2026/07}

\usepackage{tikz}
\usetikzlibrary{positioning,arrows.meta,shapes.misc,calc,decorations.pathreplacing}
\usepackage{libertine}
\usepackage{hyperref}
\usepackage{multirow}
\usepackage{hhline}
\usepackage{subcaption}
\usepackage{tcolorbox}
\usepackage[table]{xcolor}
\usepackage[overload]{textcase} 

\newcommand{\ranka}[1]{\textbf{\textcolor{black}{ #1}}}
\newcommand{\rankb}[1]{\textbf{\textcolor{darkgray}{ #1}}}
\newcommand{\rankc}[1]{\textbf{\textcolor{gray}{ #1}}}
\newcommand{\rcolor}{\cellcolor{gray!15}}
\acmSubmissionID{251}

\begin{document}

\title{Assessing Large Language Models in Generating RTL Design Specifications}


\author{Hung-Ming Huang$^{\star,1}$, Yu-Hsin Yang$^{\star,1}$, Fu-Chieh Chang$^{1}$, Yun-Chia Hsu$^{2}$, Yin-Yu Lin$^{2}$, \\ Ming-Fang Tsai$^{2}$, Chun-Chih Yang$^{2}$, Pei-Yuan Wu$^{1}$. \\
\begin{footnotesize}$^\star$Equal contribution\end{footnotesize}}
\affiliation{%
  $^1$Graduate Institute of Communication Engineering, National Taiwan University, Taipei, Taiwan \\
  $^2$MediaTek Inc, Hsinchu
  \country{Taiwan}
  }








\begin{abstract}
As IC design grows more complex, automating comprehension and documentation of RTL code has become increasingly important. Engineers currently should manually interpret existing RTL code and write specifications, a slow and error-prone process. Although LLMs have been studied for generating RTL from specifications, automated specification generation remains underexplored, largely due to the lack of reliable evaluation methods. To address this gap, we investigate how prompting strategies affect RTL-to-specification quality and introduce metrics for faithfully evaluating generated specs. We also benchmark open-source and commercial LLMs, providing a foundation for more automated and efficient specification workflows in IC design.
\end{abstract}



\maketitle

\section{Introduction}
The increasing scale and complexity of modern integrated circuit (IC) design have made Register--Transfer Level (RTL) codebases harder to document, validate, and maintain. In industrial workflows, engineers often spend considerable time manually interpreting existing RTL code to produce design specifications---documents that describe interface semantics, functional intent, timing assumptions, and control behavior. Unlike RTL, which can be simulated and verified with electronic design automation (EDA) toolchains, hardware specifications lack automated evaluation pipelines, making them labor-intensive to create and difficult to benchmark for correctness and completeness.

Recent advances in large language models (LLMs) have sparked rapid progress in \emph{Specification-to-RTL} generation (e.g.,\cite{liu2023verilogeval,pinckney2025revisiting,lu2024rtllm,liu2024openllm,pei2024betterv,10720939,11133191}), with multiple benchmarks and model improvements demonstrating strong RTL synthesis capability. However, the inverse task---generating structured, faithful specifications from RTL~\cite{li2025specllm}---remains largely under-explored, despite its critical importance in design understanding, verification hand-off, documentation, and AI-assisted hardware engineering workflows. This reverse task introduces unique challenges: unlike RTL generation, specification generation requires \emph{semantic abstraction}, \emph{temporal reasoning}, \emph{interface precision}, and \emph{preservation of design intent}, all of which are difficult to measure with traditional text- and semantic-based similarity metrics.

In this work, we present the first systematic and benchmark-driven study of RTL-to-specification generation. We argue that a correct specification should not simply restate the RTL implementation, but instead provide sufficient detail about design intent, I/O interfaces, functional behavior, state transitions, and related semantics to enable reconstruction of functionally equivalent RTL. Motivated by this, we first propose a set of prompting strategies that progressively incorporate structural guidance and reasoning steps to improve abstraction quality and reduce specification errors. Next, we introduce two hardware-aware evaluation metrics—GPT-RTL Score and RTL Reconstruction (RR) Score—that assess specification fidelity through aspect-wise analysis and testbench-level functional validation. Finally, we evaluate both commercial and open-source LLMs across multiple model scales using these metrics, providing the first quantitative comparison of model performance on the RTL-to-specification task and establishing practical recommendations for real-world deployment.  In summary, we make
the following contributions:
\begin{itemize}
    \item We present the first systematic study on the task of translating RTL code into human-readable specifications.
    \item We study how prompting strategies influence specification quality and introduce prompts tailored for RTL-to-spec generation.
    \item We propose two hardware-aware metrics that more sensitively and interpretably evaluate RTL-to-spec performance.
    \item Using these metrics, we benchmark open-source and commercial LLMs, offering guidance on model selection for practical workflows.
\end{itemize}

\section{Related Works}
\subsection{Specification-to-RTL Generation and Evaluation}
RTL generation from specifications has been widely studied. Liu et al.~\cite{liu2023verilogeval} and Pinckney et al.~\cite{pinckney2025revisiting} introduced \textit{VerilogEval} and \textit{VerilogEval-V2} for evaluating LLM-based RTL generation. Lu et al.~\cite{lu2024rtllm} and Liu et al.~\cite{liu2024openllm} proposed \textit{RTLLM} and \textit{RTLLM-2.0}, benchmarks with more complex design. Beyond benchmarks, Pei et al.~\cite{pei2024betterv} improve RTL generation via data augmentation and discriminator-guided fine-tuning. Liu et al.~\cite{10720939} develop a small-size LLM fine-tuned by RTL synthetic data outperforming commercial models. Zhao et al.~\cite{11133191} introduce a multi-agent system with sampling and debugging for RTL refinement. Despite progress in spec-to-RTL, the reverse task—RTL-to-spec generation—remains underexplored.
\subsection{RTL Code Understanding and Specification Generation}
Similar to specification generation, RTL understanding takes RTL as input and outputs natural language descriptions. Pinckney et al.~\cite{pinckney2025comprehensive} proposed \textit{CVDP}, a benchmark for RTL generation and comprehension. Liu et al.~\cite{liu2025deeprtl,liu2025deeprtl2} introduced \textit{DeepRTL} and \textit{DeepRTL2}, unified models for RTL understanding and generation evaluated via GPT Score. Although these datasets include high-level descriptions, their evaluation focuses on code comprehension rather than specification generation.
Unlike RTL understanding, specification generation requires higher-level abstraction of design intent, constraints, and functional behavior. Pinckney et al.~\cite{pinckney2025comprehensive} also note that specification generation is a more challenging test of LLM reasoning than comprehension alone. To our knowledge, the only work targeting RTL-to-spec is SpecLLM~\cite{li2025specllm}, which provides a taxonomy for specifications but relies on qualitative analysis without quantitative evaluation metrics, a gap we address in our work.

\subsection{Software Code to Document Generation and Evaluation}
Research on hardware specification generation remains limited, while the software domain has advanced rapidly in LLM-based code documentation. Shi et al.~\cite{shi2022evaluation} evaluated neural code summarization models across BLEU variants. Mastropaolo et al.~\cite{mastropaolo2024evaluating} proposed contrastive metric learning. Sun et al.~\cite{sun2024source} analyzed LLM summarization pipelines. Diggs et al.~\cite{diggs2024leveraging} studied legacy code documentation and metric correlation. Sharma et al.~\cite{sharma2025automated} introduced a structure-aware Javadoc dataset, and Fang et al.~\cite{fang2025enhanced} used a Struct-Agent for improved code interpretation. Despite progress in software documentation, applying these methods to RTL remains underexplored, as hardware specifications require modeling timing, hierarchy, and circuit behavior beyond software-style summarization.

\section{Methodology}
\subsection{Problem Formulation}
In our problem formulation, the input to the LLM is an \emph{RTL reference code} file (or repository), and the output is a \emph{functional specification} describing the design. The generated specification should be human-readable and must clearly express design intent and key hardware characteristics, including input/output interfaces, logical functionality, control flow, state transitions, and timing behaviors such as clocking and reset. Using only the generated specification, an experienced engineer should be able to reimplement RTL that is functionally equivalent to the original design.
It is important to distinguish \textit{RTL-to-specification generation} from \textit{RTL code understanding}. The core difference lies in the level of abstraction. While prior work on RTL understanding~\cite{pinckney2025comprehensive,liu2025deeprtl,liu2025deeprtl2} focuses on the  interpretation of implementaion details, specification generation~\cite{li2025specllm} requires inferring higher-level design intent and behavioral semantics. Because many functionally-equivalent RTL implementations can share the same specification, the goal of RTL-to-specification generation is not line-by-line explanation, but rather abstraction and summarization of essential logic, interfaces, and timing characteristics.

\begin{figure}[h]
    \centering
    \input{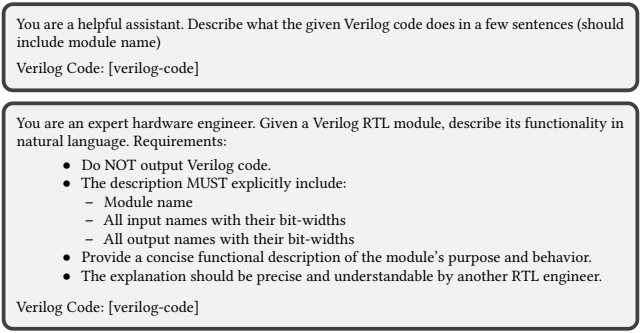}
\caption{Minimal prompt (top) and specification-aware prompt (bottom).}
\label{fig:short_and_detailed_prompts}
\end{figure}

\subsection{Prompt Design}\label{sec:prompt_design}
To generate high-quality specifications, and inspired by the prompt engineering strategies discussed in~\cite{sahoo2025systematicsurveypromptengineering}, we explore three prompt designs that provide progressively stronger structural guidance. The exact content of these three prompts are in Fig~\ref{fig:short_and_detailed_prompts} and ~\ref{fig:3stage_prompts}.
\begin{itemize}
\item \textbf{Minimal Prompt:}  
A minimal instruction that asks the model to describe the implementation of a given Verilog design in natural language. It imposes no structural or technical requirements, serving as a baseline for unconstrained specification generation.
\item \textbf{Specification-Aware Prompt:}  
A more structured prompt that explicitly instructs the model to describe essential hardware attributes, such as module interfaces (port names and bit-widths) and functional behavior. The focus is on conveying design intent and interface semantics in a clear, engineer-readable format.
\item \textbf{Multi-Step Reasoning Prompt:} 
A reasoning-enhanced prompt that enforces a staged generation process: (1) internal RTL analysis covering control logic, state behavior, and timing, (2) generation of a structured specification following a predefined template, and (3) internal self-checking to validate interface correctness and behavioral completeness. This design aims to improve consistency and reduce hallucinations.
\end{itemize}
We evaluate the effectiveness of all three prompts in Sec.~\ref{sec:exp_prompt_selection}, providing a comparative analysis to identify the prompts for producing accurate and engineer-friendly specifications.

\begin{figure}
    \centering
    \input{figures/fig_prompt_2}
\caption{Multi-step reasoning prompt.}
\label{fig:3stage_prompts}
\end{figure}

\subsection{Evaluation Metrics}\label{sec:evaluation_metric}
Prior work~\cite{li2025specllm} demonstrated RTL-to-spec generation via case studies. In contrast, we adopt a systematic, multi-metric evaluation framework inspired by software documentation research~\cite{mastropaolo2024evaluating,sun2024source}. We assess specification quality using various types of  metrics, detailed below.

\subsubsection{Textual and Semantic Similarity Metrics}
Textual similarity metrics such as BLEU~\cite{papineni2002bleu}, and ROUGE-L~\cite{lin2004rouge} measure $n$-gram overlap between generated and reference specifications, producing normalized scores in $[0,1]$. While widely used~\cite{mastropaolo2024evaluating,sun2024source,diggs2024leveraging,sharma2025automated,fang2025enhanced}, these metrics are limited in capturing semantic equivalence due to their token-level matching. Hence, semantic-based similarity metrics are proposed, such as BERTScore-R~\cite{zhang2019bertscore}, SentenceBERT Cosine Similarity (SBCS)~\cite{reimers2019sentence}. These evaluate similarity in high-dimensional semantic space, better reflecting semantic alignment~\cite{haque2022semantic}.

\subsubsection{LLM-Based Evaluation}
Following recent LLM-as-judge approaches~\cite{sun2024source,liu2025deeprtl}, we employ LLMs as evaluators, allowing them to serve as a surrogate for human assessment.

\paragraph{GPT Score}
GPT Score~\cite{liu2025deeprtl} presents both golden and generated specifications to an LLM as a judge, which outputs a similarity score within $[0,1]$ based on functional agreement in describing the Verilog design (0: no alignment, 1: identical functionality). The prompt for computing this metric is illustrated in  Fig.~\ref{fig:prompt_for_gpt_score}. We use GPT-5 Chat~\cite{openai2025gpt5} to execute this prompt, evaluating semantic and functional correctness of generated specifications.

\begin{figure}
    \centering
    \input{figures/fig_prompt_gpt_score}
    \caption{Prompt for GPT Score}
    \label{fig:prompt_for_gpt_score}
\end{figure}

\begin{figure}
    \centering
    \input{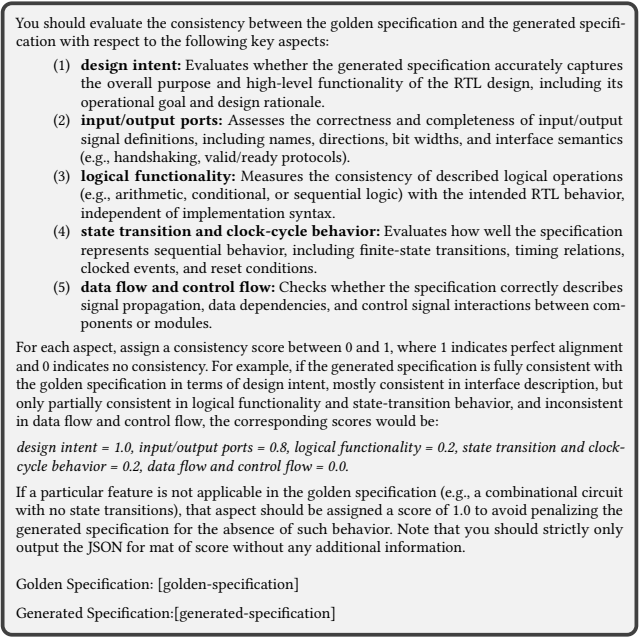}
    \caption{Prompt for GPT-RTL Score}
\label{fig:prompt_for_human_eval}
\end{figure}

\paragraph{GPT-RTL Score}
Although GPT Score primarily measures the alignment between generated and reference specifications, our goal is to make evaluation more hardware-aware. To this end, we propose GPT-RTL Score, an LLM-based assessment tailored to digital design specifications. GPT-RTL Score emphasizes hardware fidelity by evaluating five key aspects: (1) design intent, (2) input/output ports, (3) logical functionality, (4) state transition and clock-cycle behavior, and (5) data flow and control flow. The detailed prompt for evaluation is provided in Fig.~\ref{fig:prompt_for_human_eval}. For each aspect, LLMs compare consistency between golden specification and the generated specification, then assign a score between 0 and 1, where 1 indicates perfect
alignment and 0 indicates no consistency. The five dimension scores are then averaged to produce a final score between 0 and 1. By explicitly scoring these hardware-critical dimensions, GPT-RTL Score offers a more domain-aligned measure of specification quality compared to generic semantic similarity metrics. 

\subsubsection{RTL-Reconstruction Score (RR Score)}
\begin{figure}
    \centering
    \resizebox{0.48\textwidth}{!}{
    \begin{tikzpicture}[
  >=Latex,
  block/.style={draw,rounded corners,minimum width=20mm,minimum height=10mm,align=center},
  tallblock/.style={draw,rounded corners,minimum width=12mm,minimum height=25mm,align=center},
  note/.style={align=left}
]
\node[block] (spec) {Spec\\(generated)};
\node[block, above right=10mm of spec.north east, anchor=north west] (llm1) {LLM\\ (eval)};
\node[block, below right=10mm of spec.south east, anchor=south west] (llm2) {LLM \\ (oracle)};
\node[block, right=8mm of llm1] (rtl_g) {RTL\\(golden)};
\node[block, right=8mm of llm2] (rtl_gen) {RTL\\(generated)};
\node[tallblock, right=15mm of $(rtl_g)!0.5!(rtl_gen)$] (eval) { Testbench };
\node[note, right=8mm of eval.east, anchor=west] (score) {RR Score};

\draw[->] (llm1.west) -- (spec.north east);
\draw[->] (spec.south east) -- (llm2.west);
\draw[->] (rtl_g.west) -- (llm1.east) ;
\draw[->] (llm2.east) -- (rtl_gen.west);
\draw[->] (rtl_g.east) -- (eval.west |- rtl_g.east);
\draw[->] (rtl_gen.east) -- (eval.west |- rtl_gen.east);
\draw[->] (eval.east) -- (score.west);

\draw[decorate,decoration={brace,amplitude=6pt,raise=6pt}]
  ($(llm1.north west)+(-26mm,1mm)$) -- ($(rtl_g.north east)+(6mm,1mm)$)
  node[midway,above=10pt,font=\bfseries] {RTL to Spec};

\draw[decorate,decoration={brace,mirror,amplitude=6pt,raise=6pt}]
  ($(llm2.south west)+(-26mm,-1mm)$) -- ($(rtl_gen.south east)+(6mm,-1mm)$)
  node[midway,below=10pt,font=\bfseries] {Spec to RTL};

\end{tikzpicture}
    }
    \caption{RTL-Reconstruction (RR) score}
\label{fig:rtl2spec2rtl}
\end{figure}
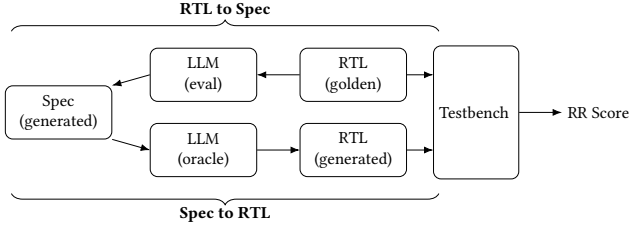

We propose the RTL-Reconstruction Score (RR Score), a behavior-grounded metric that evaluates whether a generated specification contains sufficient information—functional intent, interfaces, and timing—to reconstruct correct RTL. As shown in Fig.~\ref{fig:rtl2spec2rtl}, the top represent the task of RTL-to-specification (i,e., a reference RTL is first converted into a specification by the LLM to be evaluated). We then use an oracle LLM (GPT-5 Codex in our experiments) to reconstruct RTL based solely on the generated specification. The reconstructed and golden RTL are validated on the same testbench; behavioral equivalence determines success. The RR Score is computed as the fraction of generated specifications that pass this testbench-based equivalence check. For each specification, we use pass@1, meaning only a single reconstructed design is generated. Because correctness is measured at the input--output level, structural differences of reconstructed RTL are allowed if functionality matches. RR Score therefore assesses functional corectness rather than semantic or textual similarity. 

\section{Experiments}
\subsection{Implementation Details}
We use Python to implement our evaluation pipeline. For the LLMs being evaluated, we rely on the OpenRouter\cite{openrouter2025} API, which provides unified access to both commercial and open-source models through a single, standardized interface. This allows us to query multiple model families with consistent rate limits, formatting, and request handling. 
Details on the selected LLM candidates and evaluation benchmarks are provided below. 
\subsubsection{LLM Candidates}
We evaluate GPT-5-Codex\cite{openai_codex_chatgpt}, Claude-Sonnet-4.5\cite{anthropic_claude_sonnet}, Qwen3-Coder\cite{qwen3coder_model}, LLaMA-3.1\cite{meta_llama3.1_blog2024}, and DeepSeek-3.1\cite{deepseek2025}. For Qwen3-Coder and LLaMA-3.1, we include multiple model sizes to enable cross-scale comparison. Our study includes both commercial and open-source LLMs covering large ($>$$100$B), medium (10--100B), and small ($<$$10$B) parameter ranges, providing insight into how specification quality correlates with model scale and weight accessibility.
\subsubsection{Benchmarks}
We use two standard spec-to-RTL benchmarks, VerilogEval-V2~\cite{pinckney2025revisiting} and RTLLM-2.0~\cite{liu2024openllm}, which provide golden specifications and testbenches. Golden specs serve as the reference for assessing how closely the generated specification matches the ground truth, while testbenches verify whether generated specs enable correct RTL reconstruction. Using these benchmarks in reverse (RTL-to-spec) offers reproducible, comparable bidirectional evaluation.
\subsection{Prompt Design} 
\label{sec:exp_prompt_selection}
We evaluate the prompts described in Sec.~\ref{sec:prompt_design} using Qwen3-Coder-480B and GPT-5 Codex. The results, shown at  Table~\ref{tab:assess_llms_prompts}, highlight the best-performing prompt for each model using gray shading. Overall, the minimal prompt yields the weakest performance, while the specification-aware prompt achieves moderate improvements. The multi-step reasoning prompt produces the highest scores, demonstrating the importance of well-structured prompt design for high-quality RTL-to-specification generation. Sec.~\ref{sec:short_prompt_vs_detailed_prompt} provides an illustrative example of specifications generated using different prompts.

\begin{table}[]
\caption{Evaluation of RTL-to-Spec. 
Here, \textit{RGL} refers to \textit{ROUGE-L}, \textit{BS} denotes \textit{BertScore-R}, \textit{GS} represents \textit{GPT Score}, \textit{GRS} represents \textit{GPT-RTL Score}, \textit{RRS} denotes \textit{RTL-Reconstruction Score}.}

\begin{center}
    \begin{subtable}{0.5\textwidth}
        \centering
        \caption{Evaluation of prompts design. \textit{MINI} denotes \textit{Minimal Prompt}, \textit{SA} denotes \textit{Specification-Aware Prompt} and \textit{MSR} denotes \textit{Multi-Step Reasoning Prompt}. Gray-highlighted cells mark the best result among the three prompts for each model.}
    \resizebox{\textwidth}{!}{
    \begin{scriptsize}
    

\begin{tabular}{|p{0.7cm}|p{0.7cm}||p{0.5cm}|p{0.5cm}|p{0.5cm}|p{0.5cm}||p{0.5cm}|p{0.5cm}||p{0.5cm}|}
\hline
& Metric & \multicolumn{4}{c||}{Text \& Semantic Similarity} & \multicolumn{2}{c||}{LLM-Based} & \multicolumn{1}{c|}{RTL} \\ \hline
Model  & Prompt 
& BLEU 
& RGL 
& BS
& SBCS 
& GS 
& GRS 
& RRS 
\\ \hline\hline



\multicolumn{9}{|c|}{Benchmark: VerilogEval} \\ \hline

\multirow{3}{*}{\begin{tabular}[c]{@{}c@{}}GPT-5 \\ Codex \end{tabular}}   
&  MINI
&  2.2\%
& 20.5\%
& 81.0\%
& 64.6\%
& 90.0\% 
& {94.3\%} \rcolor{}
& 76.3\% 
\\ \hhline{~|*{8}{-}}

& SA
& {2.8\%}\rcolor{} 
& 22.6\%
& {83.0\%}
& 70.6\% \rcolor{}
& {93.5\%} \rcolor{}
& {94.2\%} 
& {89.1\%} 
\\   \hhline{~|*{8}{-}}

& MSR
&2.1\%
&29.6\% \rcolor{}
&{83.8\%} \rcolor{}
&70.5\%
& {92.4\%} 
& {94.1\%}
& {98.7\%} \rcolor{} 
\\ \hline 

\multirow{3}{*}{\begin{tabular}[c]{@{}c@{}}Qwen3\\ Coder \\ 480B\end{tabular} }   
& MINI
&  2.5\%
& 24.2\%
& 81.2\%
& 63.6\%
& 86.5\% 
& 91.4\% 
&  56.4\%
\\ \hhline{~|*{8}{-}}

& SA
& {2.7\%} \rcolor{}
& {26.3\%} \rcolor{}
& {83.0\%}
& {71.9\%} 
& {87.7\%}
& {90.5\%} 
& {73.1\%}
\\  \hhline{~|*{8}{-}}

& MSR
&2.4\%
&23.7\%
&{83.9\%} \rcolor{}
&{72.2\%} \rcolor{}
& {89.3\%} \rcolor{}
& {92.5\%} \rcolor{}
& {80.1\%} \rcolor{}
\\  \hline  \hline

\multicolumn{9}{|c|}{Benchmark: RTLLM} \\ \hline

\multirow{3}{*}{\begin{tabular}[c]{@{}c@{}}GPT-5 \\ Codex \end{tabular}}   

&  MINI
& 1.4\%
&22.6\%
&80.8\%
&82.7\%
& {94.0\%} \rcolor{}
& {95.8\%}  \rcolor{}
& 42\%
\\ \hhline{~|*{8}{-}}

&  SA
& 3.3\%
& 24.3\%
& 84.1\%
& 85.5\%
& {93.8\%} 
& {95.2\%}
& {72\%} 
\\  \hhline{~|*{8}{-}}

 &    MSR
& 4.2\% \rcolor{}
&25.8\% \rcolor{}
&86.9\% \rcolor{}
&87.8\% \rcolor{}
&{93.6\%}
& {94.8\%} 
&{76\%} \rcolor{}
\\ \hline

\multirow{3}{*}{\begin{tabular}[c]{@{}c@{}}Qwen3 \\Coder \\ 480B\end{tabular} }   
&  MINI 
&  4.4\%
& 28.2\%
& 81.7\%
& 81.4\%
& 93.4\%
& 95.0\% 
& 38\% 
\\ \hhline{~|*{8}{-}} 

&  SA  
& 6.9\%
& 29.3\%
& {85.4\%}
& 86.2\%
& {94.3\%} \rcolor{}
& {95.9\%} \rcolor{}
& {64\%} 
\\ \hhline{~|*{8}{-}}

&   MSR
& 8.0\% \rcolor{}
&{31.6\%} \rcolor{}
&{87.9\%} \rcolor{}
&90.0\% \rcolor{}
&{93.2\%}
&{95.9\%} \rcolor{}
& 66\%  \rcolor{}
\\   \hline

\end{tabular}

    \end{scriptsize}
    }
    \label{tab:assess_llms_prompts}
    \end{subtable}%
    \hfill
     \vspace{0.2cm}
    \begin{subtable}{0.5\textwidth}
    \centering
    \caption{Evaluation of LLM candidates using multi-step reasoning prompt. Bolded values indicate the top-3 scores for each metric, where black is 1st, dark gray is 2nd, and gray is 3rd.} 
    \resizebox{\textwidth}{!}{
    \begin{scriptsize}

\begin{tabular}{|p{1.7cm}||p{0.5cm}|p{0.5cm}|p{0.5cm}|p{0.5cm}||p{0.5cm}|p{0.5cm}||p{0.5cm}|}
\hline
 Metric & \multicolumn{4}{c||}{Text \& Sematic Similarity} & \multicolumn{2}{c||}{LLM-Based} & \multicolumn{1}{c|}{RTL} \\ \hline
Model
& BLEU 
& RGL 
& BS 
& SBCS 
& GS 
& GRS 
& RRS \\ \hline\hline
\multicolumn{8}{|c|}{Benchmark: VerilogEval} \\ \hline

 GPT-5-Codex 
&2.1\%
&\ranka{29.6\%} 
&\rankc{83.8\%} 
&{70.5\%}
& \ranka{92.4\%} 
& \ranka{94.1\%}
& \ranka{98.7\%} 
\\ \hline

 Claude-Sonnet-4.5 
& 1.6\%
& \rankb{27.3\%}
&82.6\%
& \ranka{72.2\%}
& \rankb{92.3\%}
& \rankb{93.4\%}
& \rankb{93.6\%}
\\ \hline

 Qwen3-Coder-480B 
&\rankb{2.4\%}
&23.7\%
&\rankb{83.9\%} 
&\ranka{72.2\%} 
& {89.3\%} 
& {92.5\%} 
& {80.1\%} 
\\ \hline

 Qwen3-Coder-30B    
& 2.3\%
& 22.9\% 
& 83.7\%
& \rankc{70.9\%}
& 85.4\%
& 89.1\% 
& 67.3\% 
\\   \hline


Llama3.1-405B   
&\ranka{2.7\%}
&\rankc{25.2\%}
&81.7\%
&69.1\%
& 84.9\%
& {89.8\%} 
& 62.2\% 
\\   \hline

 Llama3.1-70B   
&2.3\%
&{24.9\%}
&83.2\%
&67.7\%
& 84.4\%
& 90.0\% 
& 62.8\% 
\\   \hline

Llama3.1-8B   
&2.3\%
&24.2\%
&81.6\%
&68.3\%
& 77.2\%
& 84.3\% 
& 48.7\% 
\\   \hline
 
Deepseek-V3.1 
& 2.3\%
& 23.3\%
& \ranka{84.0\%}
& 70.1\%
& \rankc{91.6\%}
& \rankc{92.8\%} 
& \rankc{92.9\%}
\\  \hline\hline

\multicolumn{8}{|c|}{Benchmark: RTLLM} \\ \hline

  GPT-5-Codex 
& 4.2\% 
&25.8\% 
&86.9\% 
&87.8\% 
&\ranka{93.6\%}
&{94.8\%} 
&\ranka{76\%} 
\\ \hline

Claude-Sonnet-4.5 
&  5.5\%
& 25.1\%
& 85.9\%
& 83.7\%
& 92.7\%
& \rankb{95.5\%}
& \ranka{76\%}
\\   \hline

 Qwen3-Coder-480B   
& 8.0\% 
&\rankc{31.6\%} 
&\ranka{87.9\%} 
&90.0\% 
&\rankc{93.2\%}
&\ranka{95.9\%} 
& 66\%  
\\   \hline

 Qwen3-Coder-30B  
&\rankc{8.1\%}
&31.3\%
&\rankb{87.8\%} 
&89.7\%
& 91.8\%
& 93.5\% 
&{68\%}
\\   \hline


 Llama3.1-405B   
& \rankb{8.7\%}
&\rankb{32.8\%}
&85.2\%
&\rankb{90.3\%}
&\ranka{93.6\%}
& 93.9\% 
& 58\% 
\\   \hline

Llama3.1-70B   
&\ranka{9.3\%}
&\ranka{33.2\%}
&\rankc{87.4\%}
&\ranka{90.6\%}
& 91.7\%
& 91.4\% 
& 54\% 
\\   \hline

Llama3.1-8B   
& 8.0\%
&31.5\%
&84.8\%
&\rankc{90.1\%}
& 87.1\%
& 87.3\% 
& 46\% 
\\   \hline

Deepseek-V3.1 
& 5.7\%
&29.0\%
&87.3\%
&86.6\%
& 92.5\%
& \rankc{94.9\%} 
& \rankc{70\%}
\\  \hline
\end{tabular}

    \end{scriptsize}
    }
    \label{tab:assess_llms_candidates}
\end{subtable}

\end{center}

\end{table}

\subsection{Evaluation of LLM Candidates}

The RTL2Spec evaluation results for LLM candidates are reported at  Table~\ref{tab:assess_llms_candidates}, and the correlation between evaluation metrics are shown in Fig~\ref{fig:metric_corr_matrix}. From these results, we make the following key observations:

\begin{itemize}
    \item Overall, GPT-5-Codex delivers the best performance across most of the metrics, followed by Claude-Sonnet-4.5. Open-source models with large weights($>$$100$B), such as  DeepSeek-V3.1, Qwen-Coder-480B, and LLaMA-3.1-405B form the next tier with similar performance levels, although their relative rankings differ across datasets and evaluation metrics. For instance, under GPT Score, DeepSeek performs better on VerilogEval, while Qwen-Coder and LLaMA achieve stronger performance on RTLLM.
    \item The small-size model LLaMA-3.1-8B  performs the worst across almost all metrics except SBCS, with a significant gap from larger models. Medium-size models such as LLaMA-3.1-70B and Qwen3-Coder-30B occasionally outperform larger models on specific metrics. This suggests that aggressively reducing parameter count can substantially degrade RTL-to-spec performance, whereas well-scaled medium-size models can still retain strong reasoning and abstraction capability, offering a more efficient trade-off between model size and performance.
    \item Consistent with prior findings in~\cite{liu2025deeprtl,sun2024source}, LLM-based metrics are more reliable than token or semantic similarity metrics. Specifically, GPT Score, GPT-RTL Score and RR Score better reflect expected model behavior—latest commercial models outperform earlier open-source ones—whereas text and semantic similarity metrics often underestimate the performance gap. 
    \item Our proposed RTL-Reconstruction Score is strongly aligned with GPT Score~\cite{liu2025deeprtl}, indicating that it is also a reliable indicator of specification quality. Notably, it exhibits higher variation (46\%–98\%) than GPT Score (77\%–93\%), making it more sensitive to specification errors and more discriminative in distinguishing quality differences. A detailed comparison is provided in Sec.~\ref{sec:qa_gscore_vs_rrscore}.
    \item GPT-RTL Score is also highly correlated with GPT Score. However, unlike GPT Score, which outputs a single scalar without explanation, GPT-RTL Score evaluates five core design aspects independently before averaging them into a final score. This metric enables more interpretable assessment by identifying \emph{why} a specification receives a given score. Further analysis is presented in Sec.~\ref{sec:qa_gscore_vs_rrscore}.
    \item Although GPT-5-Codex receives a higher GPT Score than Claude-Sonnet-4.5, this result is influenced by the fact that GPT Score—and our improved variants—ultimately depend on the LLM judge in GPT series. This reliance may overestimate GPT-5-Codex’s performance while underestimating Claude-Sonnet-4.5. We discuss this limitation in Sec.~\ref{sec:limit_eval_metric}, and future work will aim to mitigate this evaluation bias.
\end{itemize}


\begin{figure}[]
\begin{center}
\begin{scriptsize}
\begin{tabular}{r c c c c c c c }
 & (1) & (2) & (3) & (4) & (5) & (6) & (7) \\
BLEU (1) &1.00\cellcolor{gray!110}  & 0.67\cellcolor{gray!77}  & 0.31\cellcolor{gray!41}  & 0.54\cellcolor{gray!64}  & 0.05\cellcolor{gray!15}  & 0.02\cellcolor{gray!12}  & -0.17\cellcolor{gray!10}  \\
ROUGE-L (2) &0.67\cellcolor{gray!77}  & 1.00\cellcolor{gray!110}  & 0.30\cellcolor{gray!40}  & 0.56\cellcolor{gray!66}  & 0.08\cellcolor{gray!18}  & 0.06\cellcolor{gray!16}  & -0.18\cellcolor{gray!10}  \\
BERTScore-R (3) &0.31\cellcolor{gray!41}  & 0.30\cellcolor{gray!40}  & 1.00\cellcolor{gray!110}  & 0.50\cellcolor{gray!60}  & 0.33\cellcolor{gray!43}  & 0.31\cellcolor{gray!41}  & 0.14\cellcolor{gray!24}  \\
SBCS (4) &0.54\cellcolor{gray!64}  & 0.56\cellcolor{gray!66}  & 0.50\cellcolor{gray!60}  & 1.00\cellcolor{gray!110}  & 0.21\cellcolor{gray!31}  & 0.18\cellcolor{gray!28}  & -0.08\cellcolor{gray!10}  \\
GPT-Score (5) &0.05\cellcolor{gray!15}  & 0.08\cellcolor{gray!18}  & 0.33\cellcolor{gray!43}  & 0.21\cellcolor{gray!31}  & 1.00\cellcolor{gray!110}  & 0.87\cellcolor{gray!97}  & 0.29\cellcolor{gray!38}  \\
GPT-RTL-Score (6) &0.02\cellcolor{gray!12}  & 0.06\cellcolor{gray!16}  & 0.31\cellcolor{gray!41}  & 0.18\cellcolor{gray!28}  & 0.87\cellcolor{gray!97}  & 1.00\cellcolor{gray!110}  & 0.28\cellcolor{gray!38}  \\
RR-Score (7) &-0.17\cellcolor{gray!10}  & -0.18\cellcolor{gray!10}  & 0.14\cellcolor{gray!24}  & -0.08\cellcolor{gray!10}  & 0.29\cellcolor{gray!38}  & 0.28\cellcolor{gray!38}  & 1.00\cellcolor{gray!110}  \\
\end{tabular}
    
\end{scriptsize}
\caption{Pairwise correlation matrix between metrics, computed using results from the Multi-step reasoning prompt across all models and datasets, where each generated specification score from a given metric is treated as an independent data point.}\label{fig:metric_corr_matrix}
\end{center}
\end{figure}

\begin{figure}
    \centering
    \input{figures/fig_sample_prompts}
\caption{An example from VerilogEval to show the comparison of results from different prompts. Certain text segments are highlighted in red for clarity, and some line breaks have been removed to meet page-length requirements.}
\label{fig:promt_example}
\end{figure}

\subsection{Qualitative Analysis}
\subsubsection{Comparison Between Different Prompts}\label{sec:short_prompt_vs_detailed_prompt}
Figure~\ref{fig:promt_example} compares outputs generated by the minimal and specification-aware prompts. The minimal prompt produces low-level, code-centric descriptions (e.g., \emph{``uses a continuous assignment to calculate the sum and carry''}) and even misidentifies the module’s functionality (e.g., \emph{``1-bit full adder''}), yielding a GPT Score of 0.6. In contrast, the specification-aware prompt guides the model toward higher-level functional intent (e.g., \emph{``calculated using basic arithmetic operation of addition''}) and produces clearer, more human-readable I/O descriptions, resulting in a higher score of 0.7.
This example demonstrates that prompt design has a substantial impact on specification quality: minimal prompts tend to produce superficial code descriptions, whereas specification-aware prompts better capture functional behavior, improve readability, and align more closely with the golden specification.

\subsubsection{Comparison between Different Metrics}\label{sec:qa_gscore_vs_rrscore}
\begin{figure}
    \centering
    \input{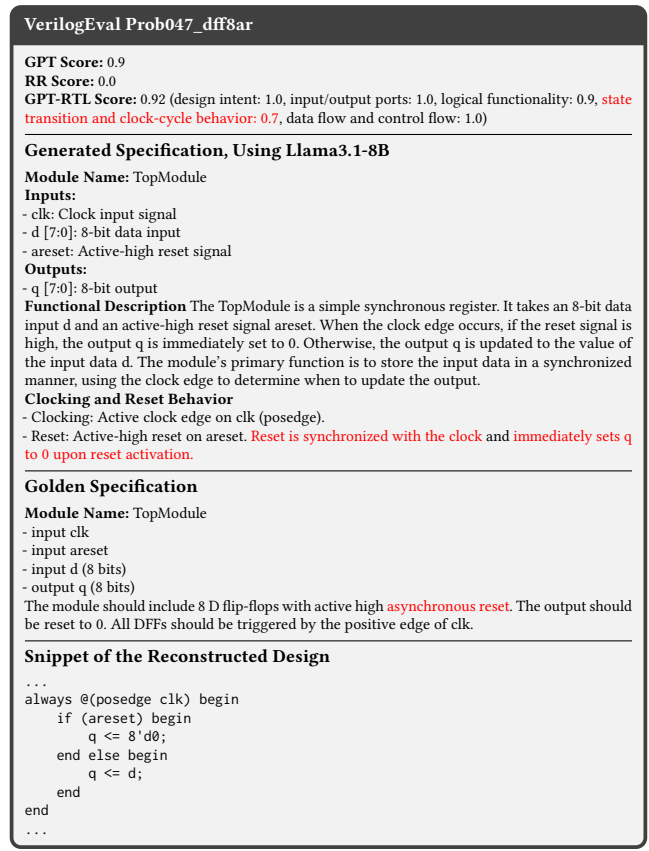}
\caption{An example from VerilogEval to show the comparison of score from different metrics. Selected text segments are highlighted in red for clarity, and some line breaks have been omitted to satisfy page-length requirements.}
\label{fig:gpt_rr_example}
\end{figure}

Fig.~\ref{fig:gpt_rr_example} shows that RTL-Reconstruction and GPT-RTL Score offer more sensitive and diagnostic evaluation than GPT Score. In this VerilogEval example, despite the golden specification requiring an \emph{asynchronous} reset, the generated specification incorrectly claims that the reset is \emph{synchronized with the clock}, but \emph{immediately sets {q} to 0 upon reset activation}. This statement contradicts both itself and the golden specification, leading the model to reconstruct an incorrect synchronous-reset design, causing a functional mismatch. Consequently, the RTL-Reconstruction Score  falls to $0.0$.
GPT-RTL Score also detects the error, reporting a score of $0.7$ for \emph{state transition and clock-cycle behavior} and consequently revealing where the inconsistency occurs. In contrast, GPT Score assigns a high score of $0.9$ without explicitly indicating the mistake. This example demonstrates that our metrics better identify subtle semantic errors and provide interpretable, aspect-level feedback.





\section{Limitations}

\subsection{Evaluation Metric}\label{sec:limit_eval_metric}
A key limitation of our work lies in the evaluation methodology. GPT Score, GPT-RTL Score, and RR Score rely on judge or oracle LLMs (GPT-5-Chat or GPT-5-Codex) to inspect specification quality or reconstruct RTL, implicitly assuming that the judge or oracle are always reliable. In practice, however, LLMs may misinterpret specifications or introduce RTL errors. For instance, GPT Score~\cite{liu2025deeprtl} uses GPT as the sole judge, which may overlook its own mistakes while more readily detecting errors produced by other models (e.g., Claude-Sonnet-4.5), potentially inflating GPT-5-Codex’s score and unfairly penalizing competing models. Human evaluation~\cite{sun2024source} can mitigate some of these issues but remains susceptible to bias and inconsistency~\cite{elangovan-etal-2024-considers}. A more promising direction is to use a panel of diverse LLMs as evaluators to reduce single-model bias, or to combine multiple LLM judges with human review. Future work should investigate such hybrid multi-LLM–human evaluation frameworks to improve robustness and trustworthiness.
\subsection{Scope of the Problems}
Our study is also limited in scope. We evaluate only single-file RTL designs from VerilogEval and RTLLM, whereas real-world designs contain multiple interacting modules. Besides, for improving RTL-to-specification generation, we focus solely on prompting and do not examine other techniques such as RAG, test-time search, or fine-tuning. Additionally, our evaluation targets functional correctness, while real specifications must express broader design objectives (e.g., PPA) and multimodal formats such as waveforms or diagrams. Extending the framework to these richer, practical settings is an important direction for future work.

\section{Conclusion}
This work presents the first systematic study of RTL-to-specification generation, addressing a critical but previously underexplored procedure of hardware design automation. We evaluate how prompting strategies influence abstraction quality, benchmark both commercial and open-source LLMs, and introduce two hardware-aware metrics—GPT-RTL Score and RTL-Reconstruction Score—that more reliably capture specification fidelity than existing measures. Our experiments on VerilogEval-V2 and RTLLM-2.0 show that structured prompting, especially the multi-step reasoning prompt, substantially improves specification correctness, while LLM-based and reconstruction-based metrics provide more sensitive and interpretable feedback on design intent, timing behavior, and interface accuracy. The evaluation further highlights clear performance differences across commercial and open-source models, with medium- and large-scale models demonstrating strong capability in generating engineer-ready specifications. We hope this study provides a foundation for future research in automated documentation and AI-assisted hardware design, enabling more accurate, scalable, and trustworthy RTL-to-specification frameworks.

\newpage
\bibliographystyle{ACM-Reference-Format}
\bibliography{reference}










\end{document}